# Languages of Dot-Depth One
# over Infinite Words [*]

Manfred Kufleitner    Alexander Lauser

University of Stuttgart, FMI

**Abstract.** Over finite words, languages of dot-depth one are expressively complete for alternation-free first-order logic. This fragment is also known as the Boolean closure of existential first-order logic. Here, the atomic formulas comprise order, successor, minimum, and maximum predicates. Knast (1983) has shown that it is decidable whether a language has dot-depth one. We extend Knast's result to infinite words. In particular, we describe the class of languages definable in alternation-free first-order logic over infinite words, and we give an effective characterization of this fragment. This characterization has two components. The first component is identical to Knast's algebraic property for finite words and the second component is a topological property, namely being a Boolean combination of Cantor sets.

As an intermediate step we consider finite and infinite words simultaneously. We then obtain the results for infinite words as well as for finite words as special cases. In particular, we give a new proof of Knast's Theorem on languages of dot-depth one over finite words.

## 1 Introduction

The investigation of logical fragments has a long history. One of the first results in our direction is due to McNaughton and Papert [22]. They showed that a language over finite words is definable in first-order logic if and only if it is star-free. A few years earlier, Schützenberger showed that a language is star-free if and only if its syntactic monoid is aperiodic [28]. For a regular language given by a (nondeterministic) finite automaton one can effectively compute its syntactic monoid and test for aperiodicity. Combining the result of McNaughton and Papert and the result of Schützenberger, this gives an algorithm for checking whether a regular language is first-order definable.

The very same approach led to similar decision procedures for various other fragments. The motivation for such results is to have some (descriptive) complexity measure for regular languages: the simpler a logical formula defining a language, the easier this language is. In addition, fragments often admit more efficient algorithms for computational problems such as the satisfiability problem. For example, the satisfiability for full first-order logic is non-elementary [30], whereas the satisfiability problem for first-order logic with only two variables is in NEXPTIME [15]. Moreover, one can frequently find temporal logic counterparts for first-order fragments and these temporal logics allow even more efficient algorithms. For example, there are temporal logics for first-order logic with two variables having a satisfiability problem in NP [9, 21]. The satisfiability problem for most temporal logics is PSPACE-complete, see e.g. [13].

---

[*]This work was supported by the German Research Foundation (DFG) under grant DI 435/5-1.

When considering some particular logical fragment $\mathcal{F}$, then there are several main aspects of $\mathcal{F}$ which are interesting: First, which languages are definable in $\mathcal{F}$, e.g., in first-order logic one can define exactly the class of star-free languages. Second, how can one decide whether a given regular language is definable in $\mathcal{F}$, e.g., a language is first-order definable if and only if its syntactic monoid is aperiodic. Third, which closure properties does $\mathcal{F}$ have, e.g., the inverse homomorphic image of a first-order definable language is again first-order definable. Other important aspects are given by relations to other fragments and the computational complexity of problems such as the satisfiability problem or the model-checking problem for $\mathcal{F}$. In this paper, we focus on the first three aspects. Very often, the second aspect is solved by giving a decidable algebraic characterization of the syntactic monoid. Apart from pure decidability, this also has the advantage that several closure properties come for free by Eilenberg's Variety Theorem [12].

The algebraic approach has been very successful for finite words [8, 33, 35, 41]. It has been generalized in different directions. One such direction is to extend the algebraic setting in order to be able to characterize more fragments. The syntactic monoid of a language and of its complement are identical. Hence, if a fragment is not closed under complementation, then only considering the syntactic monoid is not sufficient. To overcome this obstacle, Pin introduced ordered monoids and positive varieties [24]. Other fragments, such as stutter-invariant logics, are not closed under inverse homomorphisms. The solution to this problem was given by Straubing who suggested to use homomorphisms instead of semigroups or monoids. This led to the notion of $\mathcal{C}$-varieties [34, 5]. More recently Gehrke, Grigorieff, and Pin developed a general equational theory for regular languages [16].

Another way to generalize the algebraic approach is to consider other models than finite words such as infinite words [23], finite trees [3, 14], Mazurkiewicz traces [11], or data words [2], just to name a few. In most cases, considering models other than finite words requires a new notion of recognition or even new algebraic objects. The characterizations we give in this paper rely on an extended notion of recognition based on so-called linked pairs. As it turns out, purely algebraic conditions are not sufficient in this setting, but together with a topological property they work well.

When considering language classes for first-order fragments over finite words, there are two similar hierarchies within the class of star-free languages which take center stage. The first one is the dot-depth hierarchy introduced by Cohen and Brzozowski [6], and the second one is the Straubing-Thérien hierarchy [31, 36]. There is a tight connection between the two in terms of so-called wreath products [32, 40]. Both hierarchies are strict [4] and each level forms a variety [6, 26]. Thomas showed that there is a one-to-one correspondence between the quantifier alternation hierarchy of first-order logic and the dot-depth hierarchy [38]. This correspondence holds if one allows $[<, +1, \min, \max]$ as a signature. The same correspondence between the Straubing-Thérien hierarchy and the quantifier alternation hierarchy holds if we restrict the signature to $[<]$, cf. [26]. In particular, all decidability results for the dot-depth hierarchy and the Straubing-Thérien hierarchy yield decidability of the membership problem for the respective levels of the quantifier alternation hierarchy and vice versa. Unfortunately, effectively determining the level of a language in the dot-depth hierarchy or the Straubing-Thérien hierarchy is one of the most challenging open problems in automata theory. Knast has shown that the first level of the dot-depth hierarchy is decidable [20], and Simon has given a decidable characterization for the first level of the Straubing-Thérien hierarchy [29]. These two levels and the first two half levels of each hierarchy are the only decidable cases known so far, see e.g. [25] for an overview and [17] for level 3/2 of the dot-depth hierarchy. All of the above decidability results have been generalized to infinite words [1, 10, 18, 23]; the sole exception is dot-depth one. The extension of Knast's result to infinite words is the main purpose of this paper. So far, all generalizations for infinite words rely on a combination of algebraic and topological properties. As we shall see, dot-depth one is no exception.

| Fragment | Algebra | + | Topology | |
|---|---|---|---|---|
| $\Sigma_1[<]$ | $x \leq 1$ | + | Cantor sets | [23] |
| $\mathbb{B}\Sigma_1[<]$ | $\mathcal{J}$-trivial | + | Boolean combination of Cantor sets | [23] |
| $\mathbb{B}\Sigma_1[<,+1,\min]$ | $\mathbf{B_1}$ | + | Boolean combination of Cantor sets | Thm. 17 |
| $\mathrm{FO}^2[<]$ | $\mathbf{DA}$ | + | closed in strict alphabetic topology | [10] |
| $\mathrm{FO}^2[<,+1]$ | $\mathbf{LDA}$ | + | closed in strict factor topology | [18] |
| $\Sigma_2[<]$ | $eM_e e \leq e$ | + | open in alphabetic topology | [10] |
| $\Sigma_2[<,+1]$ | $eP_e e \leq e$ | + | open in factor topology | [18] |

Table 1: Fragments of first-order logic over infinite words $\Gamma^\omega$

Dot-depth one over finite words corresponds to the Boolean closure of existential first-order logic with predicates $<$ for order, $+1$ for successor, min for first position, and max for last position. This fragment is denoted by $\mathbb{B}\Sigma_1[<,+1,\min,\max]$. In our setting min and max are unary predicates rather than constants because a predicate max also makes sense for infinite words. Note that this does not change the expressive power of the fragment $\mathbb{B}\Sigma_1$ and that over infinite words the fragments $\mathbb{B}\Sigma_1[<,+1,\min]$ and $\mathbb{B}\Sigma_1[<,+1,\min,\max]$ coincide. From an algebraic and topological point of view it is more natural to work with finite and infinite words simultaneously. However, over $\Gamma^\infty = \Gamma^* \cup \Gamma^\omega$ there is one major difference between $\mathbb{B}\Sigma_1[<,+1,\min]$ without max and $\mathbb{B}\Sigma_1[<,+1,\min,\max]$ with max: The latter fragment can distinguish finite from infinite words whereas $\mathbb{B}\Sigma_1[<,+1,\min]$ cannot differentiate between $\Gamma^*$ and $\Gamma^\omega$. In particular, every $\mathbb{B}\Sigma_1[<,+1,\min]$-definable language with an infinite word also contains finite words, i.e., $\mathbb{B}\Sigma_1[<,+1,\min]$ has the finite model property.

In all variations (with or without max-predicate; infinite words $\Gamma^\omega$ only or finite and infinite words $\Gamma^\infty$) we obtain the same algebraic characterization $\mathbf{B_1}$ as Knast did for finite words. In addition, we have a topological condition which is being a finite Boolean combination of open sets. Here, *open* means open in the Cantor topology. This topological property is often denoted by $F_\sigma \cap G_\delta$, see e.g. [39]. As it turns out, there are two slightly different versions of the Cantor topology on $\Gamma^\infty$. The first one is given by base sets $u\Gamma^\infty$ for $u \in \Gamma^*$. This corresponds to the fragment $\mathbb{B}\Sigma_1[<,+1,\min]$ without max over $\Gamma^\infty$. The second version is given by base sets of the form $u\Gamma^\omega$ and $\{u\}$ for $u \in \Gamma^*$, i.e., finite words are isolated points. This second version yields a characterization of $\mathbb{B}\Sigma_1[<,+1,\min,\max]$ with max over $\Gamma^\infty$. In our setting, it is more convenient to work with some equivalent linked pair condition instead of using the topology itself.

### Related Work

Various fragments over infinite words have been considered. Existential first-order logic is denoted by $\Sigma_1$ and its Boolean closure is $\mathbb{B}\Sigma_1$. For two-variable first-order logic we write $\mathrm{FO}^2$. The second level of the alternation hierarchy is denoted by $\Sigma_2$. It contains all formulas in prenex normal form with two blocks of quantifiers, starting with a block of existential quantifiers. The prefix of a word can be defined in both $\mathrm{FO}^2[<]$ and $\Sigma_2[<]$. Hence, $\mathrm{FO}^2[<,+1,\min] = \mathrm{FO}^2[<,+1]$ and $\Sigma_2[<,+1,\min] = \Sigma_2[<,+1]$. In contrast, $\mathbb{B}\Sigma_1[<,+1]$ is a strict subclass of $\mathbb{B}\Sigma_1[<,+1,\min]$. The fragment $\Pi_2$ consists of negations of formulas in $\Sigma_2$. Since regular languages are effectively closed under complementation, decidability for $\Sigma_2$ yields decidability for $\Pi_2$.

An overview of effective characterizations can be found in Table 1. For the formal definitions of the algebraic and topological properties we refer to [10, 18, 23]. The first decision procedures for $\mathrm{FO}^2[<]$ and $\mathrm{FO}^2[<,+1]$ are due to Wilke [42], and the first effective characterization of $\Sigma_2[<]$ was given by Bojańczyk [1]. Among the topologies in Table 1, the Cantor topology is the coarsest and the strict factor topology is the finest topology. The relation between the other topologies is depicted in Figure 1.

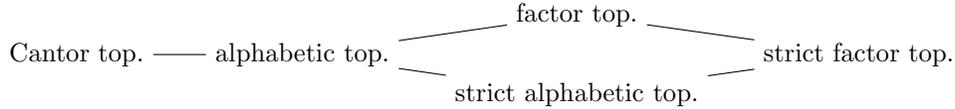

Figure 1: Topologies for infinite words.

## 2 Preliminaries

### 2.1 Languages

Throughout, $\Gamma$ is a finite nonempty alphabet. The set of finite words over $\Gamma$ is denoted by $\Gamma^*$. The empty word is $1$, and $\Gamma^+ = \Gamma^* \setminus \{1\}$ is the set of finite, nonempty words. The set of infinite words is $\Gamma^\omega$ and $\Gamma^\infty = \Gamma^* \cup \Gamma^\omega$ is the set of finite and infinite words. A *language* is a subset of $\Gamma^\infty$. Let $L \subseteq \Gamma^*$ and $K \subseteq \Gamma^\infty$. We set $LK = \{u\alpha \in \Gamma^\infty \mid u \in L, \alpha \in K\}$, $L^* = \{u_1 \cdots u_n \mid n \in \mathbb{N}, u_i \in L\}$, and $L^\omega = \{u_1 u_2 \cdots \mid u_i \in L\}$, i.e., $L^*$ is the set of finite products of words in $L$ and $L^\omega$ is the set of infinite products. We have $1^\omega = 1$. Let $\alpha \in \Gamma^\infty$ and $u \in \Gamma^*$. The word $u$ is a *factor* of $\alpha$ if $\alpha = vu\beta$ for some $v \in \Gamma^*$ and $\beta \in \Gamma^\infty$. It is a *prefix* if we can choose $v = 1$ and it is a *suffix* if we can choose $\beta = 1$. We write $u \leq \alpha$ if $u$ is a prefix of $\alpha$. The length of $\alpha$ is $|\alpha|$ and we have $|\alpha| \in \mathbb{N} \cup \{\infty\}$. For $k \in \mathbb{N}$, the *k-factor alphabet* of $\alpha$ is $\mathrm{alph}_k(\alpha) = \{u \in \Gamma^k \mid \alpha \in \Gamma^* u \Gamma^\infty\}$. If $X \subseteq \mathbb{N}$, then $\alpha(X)$ is the word comprising all positions of $\alpha$ which are contained in $X$. By extension, $\alpha(x)$ is the $x$-th letter of $\alpha$. Therefore, $\alpha = \alpha(1) \cdots \alpha(n)$ if $|\alpha| = n \in \mathbb{N}$ and $\alpha = \alpha(1)\alpha(2) \cdots$ if $|\alpha| = \infty$. We say that a position $x$ of $\alpha$ is *covered* by a factor $u$ of a factorization $\alpha = vu\beta$ if $|v| < x \leq |vu|$. If the position at which $u$ occurs in $\alpha$ is clear from the context, then we say that $u$ covers $x$. Similarly, a set of positions is covered by a set of factors if each position is covered by some factor. Here, factors are understood with implicit positions of occurrence. A *monomial* is a language of the form $w_1\Gamma^*w_2 \cdots \Gamma^*w_n$, of the form $w_1\Gamma^*w_2 \cdots \Gamma^*w_n\Gamma^\infty$, or of the form $w_1\Gamma^*w_2 \cdots \Gamma^*w_n\Gamma^\omega$ for $n \geq 1$ and $w_i \in \Gamma^*$. The *degree* of the monomial is $|w_1 \cdots w_n|$. A language $L \subseteq \Gamma^*$ of finite words has dot-depth one if it is a finite Boolean combination of monomials of the form $w_1\Gamma^*w_2 \cdots \Gamma^*w_n$. Similarly, a language $L \subseteq \Gamma^\omega$ has *dot-depth one* if it is a finite Boolean combination of monomials $w_1\Gamma^*w_2 \cdots \Gamma^*w_n\Gamma^\omega$.

## 2.2 First-Order Logic

We consider first-order logic $\text{FO} = \text{FO}[<, +1, \min, \max]$ interpreted over finite and infinite words. In the context of logic we think of words as labeled linearly ordered positions. Variables range over positions of the word. Atomic formulas are $\top$ for *true*, the unary predicates $\lambda(x) = a$, $\min(x)$ and $\max(x)$, and the binary predicates $x < y$ and $x = y+1$ for variables $x$, $y$ and $a \in \Gamma$. The formula $\lambda(x) = a$ means that $x$ is labeled with $a$, and the formula $\min(x)$ (resp. $\max(x)$) expresses that $x$ is the first (resp. last) position of the word. The formula $x < y$ is true if $x$ is strictly smaller than $y$, and $x = y + 1$ means that $x$ is the successor position of $y$. Formulas can be composed by Boolean connectives and by the quantifiers $\exists x\colon \varphi$ and $\forall x\colon \varphi$ for $\varphi \in \text{FO}$. The semantics of the connectives is as usual. A *sentence* is a formula without free variables. For a sentence $\varphi$ and for $\alpha \in \Gamma^\infty$ we write $\alpha \models \varphi$ if $\varphi$ interpreted over the word $\alpha$ is true. The *language defined* by $\varphi$ is $L(\varphi) = \{\alpha \in \Gamma^\infty \mid \alpha \models \varphi\}$.

Let $\mathcal{C} \subseteq \{<, +1, \min, \max\}$. The fragment $\Sigma_1[\mathcal{C}]$ of first-order logic consists of all formulas in FO in prenex normal form with only one block of existential quantifiers which, apart from label-predicates, use only predicates in $\mathcal{C}$. The fragment $\mathbb{B}\Sigma_1[\mathcal{C}]$ contains all finite Boolean combinations of formulas in $\Sigma_1[\mathcal{C}]$. Let $L \subseteq \Gamma^\infty$ be a language and $\mathcal{F}$ be a fragment of first-order logic. Then $L$ is *definable in* $\mathcal{F}$ if there exists some sentence $\varphi \in \mathcal{F}$ such that $L = L(\varphi)$. Sometimes we want to restrict the interpretation of the formula to some subset $K \subseteq \Gamma^\infty$. We say that $L$ is *definable in* $\mathcal{F}$ *over* $K$ if there is a sentence $\varphi \in \mathcal{F}$ with $L = \{\alpha \in K \mid \alpha \models \varphi\}$. We frequently use this with $K = \Gamma^*$ or $K = \Gamma^\omega$. Note that $\max(x)$ is false for all positions $x$ of an infinite word, i.e., a language $L$ is definable in $\mathbb{B}\Sigma_1[\mathcal{C}]$ over $\Gamma^\omega$ if and only if $L$ is definable in $\mathbb{B}\Sigma_1[\mathcal{C}, \max]$ over $\Gamma^\omega$.

## 2.3 Finite Semigroups and Finite Monoids

Let $S$ be a semigroup. An element $x \in S$ is *idempotent* if $x^2 = x$. If $S$ is finite, then there exists a number $n \geq 1$ such that the element $x^n$ is idempotent for all $x \in S$. The monoid $S^1$ generated by $S$ is defined as follows. If $S$ is a monoid, then we set $S^1 = S$; otherwise $S^1 = S \cup \{1\}$ is the monoid obtained by adding a new neutral element 1. Green's relations $\mathcal{R}$ and $\mathcal{L}$ are an important means for structural analysis in the theory of finite semigroups. For $x, y \in S$ we set

$$x \mathrel{\mathcal{R}} y \text{ iff } xS^1 = yS^1, \qquad x \leq_\mathcal{R} y \text{ iff } xS^1 \subseteq yS^1,$$
$$x \mathrel{\mathcal{L}} y \text{ iff } S^1x = S^1y, \qquad x \leq_\mathcal{L} y \text{ iff } S^1x \subseteq S^1y.$$

Remember that $xS^1 = \{xz \mid z \in S^1\}$ and $S^1x = \{zx \mid z \in S^1\}$. We often use these relations in the following way: The relation $x \leq_\mathcal{R} y$ holds if and only if there exists $z \in S^1$ such that $x = yz$. Likewise, $x \leq_\mathcal{L} y$ if and only if there exists $z \in S^1$ such that $x = zy$. As usual, we write $x <_\mathcal{R} y$ if $x \leq_\mathcal{R} y$ but not $x \, \mathcal{R} \, y$. The relation $<_\mathcal{L}$ is defined similarly.

A finite semigroup $S$ is in $\mathbf{B_1}$ if for all idempotents $e, f \in S$ and for all $s, t, x, y \in S$ we have
$$(exfy)^n exf(tesf)^n = (exfy)^n esf(tesf)^n$$
for $n \geq 1$ such that all $n$-th powers are idempotent in $S$. A semigroup $S$ is *aperiodic* if for every $x \in S$ there exists $n \geq 1$ such that $x^n = x^{n+1}$. In the equation for $\mathbf{B_1}$ we can set $e, f, s, t$ and $y$ to $x^n$ which yields $x^n x = x^n$. Hence, every semigroup in $\mathbf{B_1}$ is aperiodic. Another important property of $\mathbf{B_1}$ is given in Lemma 3 below.

The theory of first-order fragments over finite nonempty words is more concise with semigroups rather than with monoids. However, we want to treat finite and infinite words simultaneously, and our approach is heavily based on allowing the empty word 1 (and the fact that $1^\omega = 1$). On the other hand, it is crucial that the idempotents $e$ and $f$ in the above equation for $\mathbf{B_1}$ correspond to nonempty words. We therefore consider homomorphisms $h : \Gamma^* \to M$ to finite monoids. Membership in $\mathbf{B_1}$ is then formulated as $h(\Gamma^+) \in \mathbf{B_1}$.

### 2.4 Recognizability

A language $L \subseteq \Gamma^\infty$ is *regular* if it is recognized by an *extended Büchi automaton* [7], i.e., a finite automaton with two sorts of final states; the first sort is for accepting finite words and the second is for accepting infinite words by a Büchi condition. Alternatively, a language is regular if and only if it is definable in monadic second-order logic [39]. We use a more algebraic framework for recognition based on finite monoids.

Let $h : \Gamma^* \to M$ be a homomorphism to a finite monoid $M$. If $h$ is understood and $s \in M$, then we write $[s]$ for the language $h^{-1}(s)$. A *linked pair* of $M$ is a pair $(s, e) \in M \times M$ such that $e$ is idempotent and $s = se$. For every word $\alpha \in \Gamma^\infty$ there exists a linked pair $(s, e)$ of $M$ such that $\alpha \in [s][e]^\omega$ by Ramsey's Theorem [27]. A language $L \subseteq \Gamma^\infty$ is *recognized* by $h$ if
$$L = \bigcup \left\{ [s][e]^\omega \mid (s, e) \text{ is a linked pair with } [s][e]^\omega \cap L \neq \emptyset \right\}.$$

The *syntactic congruence* of $L \subseteq \Gamma^\infty$ is defined as follows. For nonempty words $p, q \in \Gamma^+$ we let $p \equiv_L q$ if for all words $u, v, w \in \Gamma^*$ the following equivalences hold:
$$upvw^\omega \in L \Leftrightarrow uqvw^\omega \in L \quad \text{and}$$
$$u(pv)^\omega \in L \Leftrightarrow u(qv)^\omega \in L.$$

Remember that $1^\omega = 1$. This relation indeed is a congruence and the congruence classes $[p]_L = \{q \in \Gamma^+ \mid p \equiv_L q\}$ constitute the *syntactic semigroup* $\mathrm{Synt}(L)$. The *syntactic monoid* $\mathrm{Synt}^1(L)$ is the monoid generated by $\mathrm{Synt}(L)$, i.e., $\mathrm{Synt}^1(L) = S^1$ for $S = \mathrm{Synt}(L)$. The *syntactic homomorphism* $h_L : \Gamma^* \to \mathrm{Synt}^1(L)$ is defined by $h_L(a) = [a]_L$ for $a \in \Gamma$. A variant of the syntactic monoid is the *pure syntactic monoid* $\mathrm{Synt}_+(L) = \mathrm{Synt}(L) \dot\cup \{1\}$, i.e., we add a new neutral element to $\mathrm{Synt}(L)$, even if $\mathrm{Synt}(L)$ is a monoid. The *pure syntactic homomorphism* $h_+ : \Gamma^* \to \mathrm{Synt}_+(L)$ is defined by $h_+(p) = h_L(p)$ for $p \neq 1$. The only possible difference between $h_+$ and $h_L$ is their behavior on the empty word. Note that
$$h_L(\Gamma^+) = h_+(\Gamma^+) = \mathrm{Synt}(L) \subseteq \mathrm{Synt}^1(L) \subseteq \mathrm{Synt}_+(L)$$

and $\text{Synt}_+(L) \setminus \{1\} = \text{Synt}(L) \subsetneq \text{Synt}_+(L)$. A language $L \subseteq \Gamma^\infty$ is regular if and only if both $\text{Synt}(L)$ is finite and $h_L$ recognizes $L$, see e.g. [23, 39]. Moreover, $L$ is recognized by its syntactic homomorphism $h_L$ if and only if it is recognized by its pure syntactic homomorphism $h_+$. In contrast to $h_L$, the pure syntactic homomorphism has the property that $h_+(u) = 1$ if and only if $u = 1$.

**Lemma 1.** *Let $L \subseteq \Gamma^\infty$ be recognized by a homomorphism $h : \Gamma^* \to M$ such that $h(u) = 1$ only if $u = 1$. Then both $L \cap \Gamma^*$ and $L \cap \Gamma^\omega$ are also recognized by $h$.*

*Proof:* We have $[s] = [s][1]^\omega \subseteq \Gamma^*$. Moreover, $[s][e]^\omega \subseteq \Gamma^\omega$ if $e \neq 1$. This proves the claim. □

## 3 Algebraic Properties

This section contains simple algebraic and combinatorial properties of the class $\mathbf{B_1}$. The following elementary lemma gives a mechanism for obtaining idempotent stabilizers with a nonempty preimage: Every sufficiently long word $u$ has a short prefix $p$ admitting a nonempty idempotent stabilizer $e$.

**Lemma 2.** *Let $h : \Gamma^* \to M$ be a homomorphism to a finite monoid $M$ and let $u \in \Gamma^*$ with $|u| = |M| - 1$. Then there exists a prefix $p$ of $u$ and an idempotent $e \in h(\Gamma^+)$ with $h(p)e = h(p)$.*

*Proof:* Let $a \in \Gamma$ and let $1 = p_0 < p_1 < \cdots < p_{|M|} = ua$ be the prefixes of $ua$. By the pigeonhole principle, there exist $0 \leq i < j \leq |M|$ such that $h(p_i) = h(p_j)$. In particular, we have $i \leq |M| - 1$ and $p_i$ is a prefix of $u$. Let $p_i q = p_j$ for $q \in \Gamma^+$. We set $e = h(q)^n$ to be the idempotent element generated by $h(q)$. Now, $h(p)e = h(p)$ for $p = p_i$. □

Next we state the key property of $\mathbf{B_1}$, a substitution rule valid in certain situations. Much of the work in proving our main theorem is devoted to guarantee its premises.

**Lemma 3.** *Let $S \in \mathbf{B_1}$. If $u \mathcal{R} uexf$ and $esfv \mathcal{L} v$ for idempotents $e, f \in S$ and for $u, v, x, s \in S$, then $uexfv = uesfv$.*

*Proof:* Choose $n \geq 1$ such that all $n$-th powers in $S$ are idempotent. Since $u \mathcal{R} uexf$ and $v \mathcal{L} esfv$, there exist $y, t \in S^1$ with $u = uexfy$ and $v = tesfv$. In particular, $u = u(exfy)^n$ and $v = (tesf)^n v$. We can assume $y, t \in S$ because $e$ and $f$ are idempotent. Using the equation for $\mathbf{B_1}$ we conclude

$$uexfv = u(exfy)^n exf(tesf)^n v$$
$$= u(exfy)^n esf(tesf)^n v = uesfv. \qquad \Box$$

Proposition 4 below gives an important combinatorial feature of $\mathbf{B_1}$. It shows that if the $\mathcal{R}$-class changes when reading a word from left to right (resp. the $\mathcal{L}$-class changes when reading the word from right to left), then this happens with a new factor of bounded length.

**Proposition 4.** *Let $h : \Gamma^* \to M$ be a homomorphism with $h(\Gamma^+) \in \mathbf{B_1}$ and let $k \geq |M|$. For all $a \in \Gamma$ and $u, x \in \Gamma^*$ with $|x| \geq k$ we have:*
  1. $h(u) \mathcal{R} h(ux) >_\mathcal{R} h(uxa) \Rightarrow \text{alph}_k(x) \neq \text{alph}_k(xa)$.
  2. $h(u) \mathcal{L} h(xu) >_\mathcal{L} h(axu) \Rightarrow \text{alph}_k(x) \neq \text{alph}_k(ax)$.

*Proof:* By left-right symmetry, it suffices to show "1". Assume $h(u) \mathcal{R} h(ux) >_{\mathcal{R}} h(uxa)$ and $\mathrm{alph}_k(x) = \mathrm{alph}_k(xa)$. Let $w$ be the suffix of length $k$ of $xa$. By Lemma 2, there exist $y, z \in \Gamma^*$ with $w = yza$ and $h(y)e = h(y)$ for some idempotent $e \in h(\Gamma^+)$ because $|w| \geq |M|$. Let $|y|$ be maximal with this property. Since $w \in \mathrm{alph}_k(xa) = \mathrm{alph}_k(x)$, we can write $x = syzatz$ for some $s, t \in \Gamma^*$ such that $y$ is a suffix of $yzat$. Note that there is indeed at least one letter between the two occurrences of $z$. Let $u' = h(usy)$ and $x' = h(zat)$. We have $u' = u'e$, $u'x' = u'x'e$, and there exists $y' \in h(\Gamma^+)$ with $u' = u'x'y'$. Therefore, we have $u'x' = u'(ex'ey')^n ex'e(eeee)^n$ for all $n \in \mathbb{N}$, and by $h(\Gamma^+) \in \mathbf{B_1}$ this equals $u'(ex'ey')^n eee(eeee)^n = u'$ for sufficiently large $n$. Thus $u' = u'x'$ and $h(u) \mathcal{R} u' = u'x'x' \mathcal{R} h(uxa)$, contradicting the assumption. □

## 4 The Fragment $\mathbb{B}\Sigma_1[<, +1, \min]$ over $\Gamma^\infty$

This section contains our main result Theorem 5. We give an effective characterization of the first-order fragment $\mathbb{B}\Sigma_1[<, +1, \min]$ over finite and infinite words.

Over $\Gamma^\infty$, the fragment $\mathbb{B}\Sigma_1[<, +1, \min]$ yields a strict subclass of the $\mathbb{B}\Sigma_1[<, +1, \min, \max]$-definable languages. For example, $\Gamma^*a$ is not definable in alternation-free first-order logic without max-predicate. On the other hand, the language $a\Gamma^\infty$ is definable in $\mathbb{B}\Sigma_1[\min]$. We pinpoint this asymmetry of $\mathbb{B}\Sigma_1[<, +1, \min]$ to some topological condition (expressed in terms of linked pairs).

**Theorem 5.** *Let $L \subseteq \Gamma^\infty$ be regular. The following assertions are equivalent:*
1. *$L$ is a finite Boolean combination of monomials of the form $w_1\Gamma^*w_2 \cdots \Gamma^*w_n\Gamma^\infty$.*
2. *$L$ is definable in $\mathbb{B}\Sigma_1[<, +1, \min]$.*
3. *The syntactic homomorphism $h_L : \Gamma^* \to \mathrm{Synt}^1(L)$ satisfies*
   *a) $\mathrm{Synt}(L) \in \mathbf{B_1}$, and*
   *b) for all linked pairs $(s, e)$ and $(t, f)$ of $\mathrm{Synt}^1(L)$ with $s \mathcal{R} t$ we have $[s][e]^\omega \subseteq L \Leftrightarrow [t][f]^\omega \subseteq L$.*
4. *$L$ is recognized by a homomorphism $h : \Gamma^* \to M$ satisfying*
   *a) $h(\Gamma^+) \in \mathbf{B_1}$, and*
   *b) for all linked pairs $(s, e)$ and $(t, f)$ of $M$ with $s \mathcal{R} t$ we have $[s][e]^\omega \subseteq L \Leftrightarrow [t][f]^\omega \subseteq L$.*

**Remark 6.** *Suppose $h : \Gamma^* \to M$ recognizes a regular language $L$ and consider the condition $[s][e]^\omega \subseteq L \Leftrightarrow [t][f]^\omega \subseteq L$ for all linked pairs $(s, e)$ and $(t, f)$ of $M$ with $s \mathcal{R} t$. This condition is equivalent to $L$ being a finite Boolean combination of open sets, cf. [23, Theorem VI.3.7]. Here, open means open in the Cantor topology defined by the base sets $u\Gamma^\infty$ for $u \in \Gamma^*$. Therefore, the conditions "3b" and "4b" in Theorem 5 are actually topological properties.*

**Remark 7.** *For languages over $\Gamma^\infty$ there is also the concept of weak recognition. A language $L$ is weakly recognized by a homomorphism $h : \Gamma^* \to M$ to a finite monoid if*

$$L = \bigcup \{[s][e]^\omega \mid (s, e) \text{ is a linked pair with } [s][e]^\omega \subseteq L\}.$$

*If a language $L \subseteq \Gamma^\infty$ is recognized by a homomorphism $h$, then it is weakly recognized by $h$. In general the converse is not true. However, if in addition $[s][e]^\omega \subseteq L \Leftrightarrow [t][f]^\omega \subseteq L$ for all linked pairs $(s, e)$ and $(t, f)$ of $M$ with $s \mathcal{R} t$, then weak recognition implies strong recognition. Suppose $[s][e]^\omega \cap L \neq \emptyset$. Then there exists a linked pair $(t, f)$ with $[t][f]^\omega \subseteq L$ and $[s][e]^\omega \cap [t][f]^\omega \neq \emptyset$. The latter condition implies $s \mathcal{R} t$ and hence $[s][e]^\omega \subseteq L$.*

In the remainder of this section we prove Theorem 5. The implications "1 ⇒ 2" and "2 ⇒ 3" are Lemmas 8, 9 and 10. The most involved part "4 ⇒ 1" is shown in the second half of this section.

**Lemma 8.** *Let $n \geq 1$ and let $w_1, \ldots, w_n \in \Gamma^*$.*

1. *The monomial $w_1 \Gamma^* w_2 \cdots \Gamma^* w_n \Gamma^\infty$ is defined by a sentence in $\Sigma_1[<, +1, \min]$ with quantifier depth $|w_1 \cdots w_n|$.*
2. *The monomial $w_1 \Gamma^* w_2 \cdots \Gamma^* w_n$ is defined by a sentence in $\Sigma_1[<, +1, \min, \max]$ with quantifier depth $|w_1 \cdots w_n|$.*

*Proof:* We write $\equiv$ for syntactic equivalence of formulas. For variable vectors $\underline{x} = (x_1, \ldots, x_\ell)$ and $\underline{y} = (y_1, \ldots, y_m)$ we introduce the shortcuts $\exists \underline{x} \equiv \exists x_1 \cdots \exists x_\ell$, $\min(\underline{x}) \equiv \min(x_1)$, $\max(\underline{x}) \equiv \max(x_\ell)$, $\underline{x} < \underline{y} \equiv x_\ell < y_1$, and $\lambda(\underline{x}) = a_1 \cdots a_\ell$ for

$$\bigwedge_{1 \leq j \leq \ell} \lambda(x_j) = a_j \wedge \bigwedge_{1 \leq j < \ell} x_{j+1} = x_j + 1.$$

Let $L = w_1 \Gamma^* w_2 \cdots \Gamma^* w_n \Gamma^\infty$. We introduce variable vectors $\underline{x}_i = (x_{i,1}, \ldots, x_{i,|w_i|})$ for every $i \in \{1, \ldots, n\}$. Then $L$ is defined by the following sentence $\varphi$:

$$\exists \underline{x}_1 \cdots \exists \underline{x}_n \colon \min(\underline{x}_1) \wedge \bigwedge_{1 \leq i \leq n} \lambda(\underline{x}_i) = w_i \wedge \bigwedge_{1 \leq i < n} \underline{x}_i < \underline{x}_{i+1}.$$

The second term of the conjunction ensures that each $\underline{x}_i$ corresponds to a factor $w_i$ and the first term says that any model starts with $w_1$. The third term makes sure that the factors $w_i$ occur in the correct order. The sentence for $w_1 \Gamma^* w_2 \cdots \Gamma^* w_n$ is $\varphi \wedge \max(\underline{x}_n)$. □

**Lemma 9.** *If $L \subseteq \Gamma^\infty$ is definable in $\mathbb{B}\Sigma_1[<, +1, \min, \max]$, then $\mathrm{Synt}(L) \in \mathbf{B_1}$.*

*Proof:* Let $e, f, s, t, x, y \in \Gamma^+$, let $n \geq 1$ and define

$$p = (e^n x f^n y)^n e^n x f^n (t e^n s f^n)^n,$$
$$q = (e^n x f^n y)^n e^n s f^n (t e^n s f^n)^n.$$

For all $u, v, w \in \Gamma^*$ and for all sentences $\varphi \in \Sigma_1[<, +1, \min, \max]$ with quantifier depth at most $n$, we show $upvw^\omega \models \varphi$ if and only if $uqvw^\omega \models \varphi$. Let $\psi$ be quantifier free such that $\varphi = \exists x_1 \cdots \exists x_n \colon \psi$. Suppose $upvw^\omega \models \varphi$ and consider positions $x_i$ such that $\psi$ is true. The consecutive positions in this assignment induce a sequence of factors $w_1, \ldots, w_m$ of $upvw^\omega$ with $m \leq n$ and $|w_i| \leq n$ for all $i$. Since this sequence of nonadjacent factors appears in the same order in $uqvw^\omega$, we see that $uqvw^\omega \models \varphi$. Showing that $uqvw^\omega \models \varphi$ implies $upvw^\omega \models \varphi$ is symmetric.

The equivalence of $u(pv)^\omega \models \varphi$ and $u(qv)^\omega \models \varphi$ is similar. Thus the syntactic semigroup of every $\mathbb{B}\Sigma_1[<, +1, \min, \max]$-definable language is in $\mathbf{B_1}$. □

**Lemma 10.** *Let $L \subseteq \Gamma^\infty$ be definable in $\mathbb{B}\Sigma_1[<, +1, \min]$ and let $M$ be a finite monoid. For every surjective homomorphism $h \colon \Gamma^* \to M$ which recognizes $L$ we have $[s] \subseteq L \Leftrightarrow [s][e]^\omega \subseteq L$ for every linked pair $(s, e)$ of $M$.*

*Proof:* Let $\varphi \in \Sigma_1[<, +1, \min]$ be a sentence. If $\alpha \in \Gamma^\infty$ models $\varphi$, then there is a prefix $u$ of $\alpha$ such that for every $\beta \in \Gamma^\infty$ we have $u\beta \models \varphi$. This is because $\alpha \models \varphi$ yields some satisfying assignment for the variables, and positions beyond the last position of this assignment have no influence. Let $L$ be defined by a sentence in $\mathbb{B}\Sigma_1[<, +1, \min]$ with quantifier depth $d$. Consider $\alpha = \hat{s}\,\hat{e}^\omega$ for $\hat{s} \in [s]$ and $\hat{e} \in [e]$. By the above consideration there exists a finite prefix $u = \hat{s}\,\hat{e}^n$ of $\alpha$ such that $\alpha$ and $u$ model the same formulas in $\Sigma_1[<, +1, \min]$ with quantifier depth at most $d$. Now, $u \in L$ if and only if $\alpha \in L$. Therefore, $[s] \subseteq L$ if and only if $[s][e]^\omega \subseteq L$. □

In the remainder of this section we show that condition "4" in Theorem 5 is sufficient. To this end we show that if $\alpha$ and $\beta$ are contained in the same monomials up to a certain degree, then their images in a semigroup in $\mathbf{B_1}$ are $\mathcal{R}$-related. The main idea is to apply Lemma 3. The first step is to show that under certain conditions we can replace several factors in finite words (Lemma 11). To formulate these conditions we introduce the $\mathcal{R}(k)$-factorization and the $\mathcal{L}(k)$-factorization. Then the substitution principle in Lemma 11 is extended to infinite words (Lemma 12). Finally, in Proposition 13 we show that in $\mathbf{B_1}$ we can guarantee the premises of Lemma 12.

We think of a factor $u_i$ as being equipped with the position $x_i$ of its first letter. Consequently, a *factorization* $F$ is a tuple $(x_1, u_1, \ldots, x_\ell, u_\ell) \in (\mathbb{N} \times \Gamma^+)^\ell$ with $\ell \geq 0$ and $x_{i+1} \geq x_i + |u_i|$ for all $1 \leq i < \ell$, i.e., we assume that the factors $u_i$ are in increasing order and nonoverlapping. The *type* of $F$ is the sequence of words $(u_1, \ldots, u_\ell)$. We say that $F$ is a *factorization of $\alpha$* if $u_i = \alpha(\{x_i, \ldots, x_i + |u_i| - 1\})$ for all $1 \leq i \leq \ell$.

We want to merge two factorizations $F = (x_1, u_1, \ldots, x_\ell, u_\ell)$ and $G = (y_1, v_1, \ldots, y_m, v_m)$ of $\alpha$. In order to define the *join* $F \vee G$ of $F$ and $G$, we combine overlapping factors of $F$ and $G$ into one factor, see Figure 2 for an illustration. More precisely, let $X_i = \{x_i, \ldots, x_i + |u_i| - 1\}$ be the positions of the factor $u_i$ and let $Y_i = \{y_i, \ldots, y_i + |v_i| - 1\}$ be the positions of the factor $v_i$. We say that $X = \bigcup_{i=1}^\ell X_i$ is the *set of positions of $F$*. Analogously, $Y = \bigcup_{i=1}^m Y_i$ is the set of positions of $G$. We set $Z = X \cup Y$. Let $\{Z_1, \ldots, Z_n\}$ be the finest partition of $Z$ such that every class $Z_j$ is a union of sets $X_i$ and sets $Y_i$. Therefore, if $x < y < z$ and $x, z \in Z_j$, then $y \in Z_j$; otherwise we could split $Z_j$ into two classes $Z_j \cap \{s \in \mathbb{N} \mid s < y\}$ and $Z_j \cap \{s \in \mathbb{N} \mid s > y\}$, resulting in a finer partition. Therefore, each $\alpha(Z_j)$ is a factor of $\alpha$. Let $z_j$ be the minimal element in $Z_j$ and suppose $z_1 < \cdots < z_n$. Now, the join of $F$ and $G$ is

$$F \vee G = \big(z_1, \alpha(Z_1), \ldots, z_n, \alpha(Z_n)\big).$$

It is easy to see that the operation $\vee$ on factorizations of $\alpha$ is associative and commutative.

An important algebraic concept in our proofs is the $\mathcal{R}(k)$-factorization and its left-right dual, the $\mathcal{L}(k)$-factorization. Let $h : \Gamma^* \to M$ be a homomorphism to a finite monoid $M$. The $\mathcal{R}$-*factorization* of a word $\alpha$ is given by the positions where the $\mathcal{R}$-class changes when reading $\alpha$ from left to right. More precisely, let $\alpha = a_1 w_1 \cdots a_{r-1} w_{r-1} a_r \beta$ with $r \geq 0$, $a_i \in \Gamma$, $w_i \in \Gamma^*$ and $\beta \in \Gamma^\infty$ such that

$$h(a_1 w_1 \cdots a_i) \mathrel{\mathcal{R}} h(a_1 w_1 \cdots a_i w_i) >_\mathcal{R} h(a_1 w_1 \cdots a_{i+1})$$

for all $1 \leq i < r$ and $h(a_1 w_1 \cdots a_r) \mathrel{\mathcal{R}} h(a_1 w_1 \cdots a_r w)$ for every finite prefix $w$ of $\beta$. Let $z_i$ be the position of $a_i$ in the above factorization. The $\mathcal{R}$-factorization of $\alpha$ is $(z_1, a_1, \ldots, z_r, a_r)$. For every word $\alpha$, the above factorization is unique and its size $r$ is at most $|M|$. Note that $x_1 = 1$ for every nonempty word $\alpha$, even if $h(a_1) = 1$.

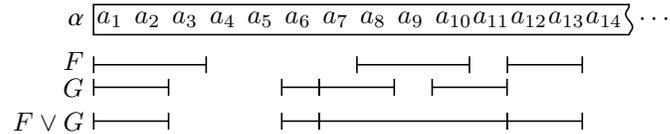

Figure 2: The join $F \vee G$ of the factorizations $F$ and $G$ obtained by merging overlapping factors. Here, we have $F = (1, a_1a_2a_3, 8, a_8a_9a_{10}, 12, a_{12}a_{13})$ and $G = (1, a_1a_2, 6, a_6, 7, a_7a_8, 10, a_{10}a_{11})$. The join of these two factorizations is $F \vee G = (1, a_1a_2a_3, 6, a_6, 7, a_7a_8a_9a_{10}a_{11}, 12, a_{12}a_{13})$. Note that nonoverlapping adjacent factors are not merged.

We extend this definition by taking the contexts of the $\mathcal{R}$-factorization into account. Let $k \in \mathbb{N}$ and consider the $\mathcal{R}$-factorization $(z_1, a_1, \ldots, z_r, a_r)$ of $\alpha$. Let $F_i = (z'_i, w_i)$ with $z'_i = \max\{1, z_i - k\}$ and $w_i = \alpha(\{z_i - k, \ldots, z_i + k\})$, i.e., $w_i$ is the factor of $\alpha$ induced by all positions $z$ such that $|z - z_i| \leq k$. The $\mathcal{R}(k)$-*factorization* of $\alpha$ is $F_1 \vee \cdots \vee F_r$. Let $F = (x_1, u_1, \ldots, x_\ell, u_\ell)$ be the $\mathcal{R}(k)$-factorization of $\alpha$ and let $X$ be the set of its positions. We have $|X| \leq |M|(2k+1) - k$ since at most $k+1$ positions come from the first position of the $\mathcal{R}$-factorization and all other positions of the $\mathcal{R}$-factorization contribute at most $2k+1$ positions to $X$. In particular, $|X| \leq 2k^2$ if $k \geq |M|$. We have $\alpha = u_1 w_1 \cdots u_{\ell-1} w_{\ell-1} u_\ell \beta$ for some $w_i \in \Gamma^*$, $\beta \in \Gamma^\infty$ such that the $u_i$'s cover the positions of the $\mathcal{R}$-factorization and moreover, the $\mathcal{R}$-class changes at neither the $k$ first positions of any $u_i$ with $i > 1$ nor at the $k$ last positions of any $u_i$ with $i < \ell$.

The $\mathcal{L}$-factorization of a finite word $w \in \Gamma^*$ is the left-right dual of the $\mathcal{R}$-factorization: Let $w = w_1 a_1 \cdots w_r a_r$ with $r \geq 0$, $a_i \in \Gamma$, and $w_i \in \Gamma^*$ such that

$$h(a_{i-1} w_i a_i \cdots w_r a_r) <_\mathcal{L} h(w_i a_i \cdots w_r a_r) \mathrel{\mathcal{L}} h(a_i \cdots w_r a_r)$$

for all $1 \leq i \leq r$. The $\mathcal{L}$-*factorization* of $w$ is then given by the factors $a_i$ of length one together with their positions in $w$.

As for $\mathcal{R}$-factorizations, we extend this definition by taking contexts into account. Let $(z_1, a_1, \ldots, z_r, a_r)$ be the $\mathcal{L}$-factorization of $w$. Let $k \in \mathbb{N}$ and let $G_i = (z'_i, w_i)$ with $z'_i = \max\{1, z_i - k\}$ and let $w_i = w(\{z_i - k, \ldots, z_i + k\})$ be the factor of $w$ induced by all positions $z$ such that $|z - z_i| \leq k$. Then, the $\mathcal{L}(k)$-*factorization* of $w$ is $G_1 \vee \cdots \vee G_r$. Let $G = (y_1, v_1, \ldots, y_m, v_m)$ be the $\mathcal{L}(k)$-factorization of $w$ and let $Y$ be the set of its positions. As for $\mathcal{R}(k)$-factorizations, we have $|Y| \leq 2k^2$ if $k \geq |M|$.

**Lemma 11.** *Let* $h : \Gamma^* \to M$ *with* $h(\Gamma^+) \in \mathbf{B_1}$ *and let* $k \geq |M|$. *If* $u = w_0 u_1 w_1 \cdots u_\ell w_\ell$ *and* $v = w_0 v_1 w_1 \cdots v_\ell w_\ell$ *for words* $u_i, v_i, w_i \in \Gamma^*$ *such that the* $w_i$'s *in* $u$ *cover the positions of the* $\mathcal{R}(k)$-*factorization of* $u$ *and the* $w_i$'s *in* $v$ *cover the positions of the* $\mathcal{L}(k)$-*factorization of* $v$, *then* $h(u) = h(v)$.

*Proof:* The proof goes as follows. Since $k$ is large enough, we find a short prefix $p_i$ and a short suffix $q_i$ of each $w_i$ admitting idempotent stabilizers $f_i$ and $e_i$. Appending these prefixes and suffixes to the $u_i$'s and $v_i$'s then allows us to apply Lemma 3.

We can assume that each $w_i$ covers the positions of a factor of the $\mathcal{R}(k)$-factorization of $u$ or of a factor of the $\mathcal{L}(k)$-factorization of $v$. In particular, $|w_0|, |w_\ell| \geq k$ and $|w_i| \geq 2k$ for $0 < i < \ell$. By Lemma 2 and its left-right dual, there exist idempotents $f_1, \ldots, f_\ell, e_0, \ldots, e_{\ell-1} \in h(\Gamma^+)$ such that each $w_i$ admits a factorization $w_i = p_i r_i q_i$ with $|p_i| \leq k$ and $|q_i| \leq k$ satisfying

$$h(p_i) = h(p_i) f_i \quad \text{for } 0 < i \leq \ell,$$
$$h(q_i) = e_i h(q_i) \quad \text{for } 0 \leq i < \ell.$$

In particular, we can assume $p_0 = 1 = q_\ell$. Let $x_i = q_{i-1} u_i p_i$ and $s_i = q_{i-1} v_i p_i$ for $1 \leq i \leq \ell$. Then, $u = r_0 x_1 r_1 \cdots x_\ell r_\ell$ and $s = r_0 s_1 r_1 \cdots s_\ell r_\ell$, and the $r_i$'s in $u$ cover the positions of the $\mathcal{R}$-factorization of $u$, whereas the $r_i$'s in $v$ cover the positions of the $\mathcal{L}$-factorization of $v$. Thus

$$h(r_0 x_1 \cdots r_{i-1}) \mathcal{R} \, h(r_0 x_1 \cdots r_{i-1}) \cdot e_{i-1} h(x_i) f_i,$$
$$e_{i-1} h(s_i) f_i \cdot h(r_i \cdots s_\ell r_\ell) \mathcal{L} \, h(r_i \cdots s_\ell r_\ell).$$

An $\ell$-fold application of Lemma 3 yields

$$\begin{aligned} h(u) &= h(r_0 x_1 \cdots r_{\ell-2} x_{\ell-1} r_{\ell-1} x_\ell r_\ell) \\ &= h(r_0 x_1 \cdots r_{\ell-2} x_{\ell-1} r_{\ell-1} s_\ell r_\ell) \\ &= h(r_0 x_1 \cdots r_{\ell-2} s_{\ell-1} r_{\ell-1} s_\ell r_\ell) \\ &= \cdots \\ &= h(r_0 s_1 \cdots r_{\ell-2} s_{\ell-1} r_{\ell-1} s_\ell r_\ell) = h(v). \end{aligned}$$

We can think of the above equations as converting $h(u)$ into $h(v)$ by using substitution rules $x_i \to s_i$. Note that the image under $h$ is preserved only when applying these rules from right to left. □

Next, we give a version of Lemma 11 for finite and infinite words. The problem is that there is no canonical choice for the $\mathcal{L}(k)$-factorization of an infinite word $\alpha$. We overcome this obstacle by fixing a type and considering $\mathcal{L}(k)$-factorizations of this type for infinitely many prefixes of $\alpha$.

**Lemma 12.** *Let $h : \Gamma^* \to M$ with $h(\Gamma^+) \in \mathbf{B_1}$, let $k \geq |M|$ and let $\alpha = w_0 u_1 w_1 \cdots u_\ell w_\ell \gamma$ with $u_i, w_i \in \Gamma^*$ and $\gamma \in \Gamma^\infty$ such that the $w_i$'s cover the positions of the $\mathcal{R}(k)$-factorization of $\alpha$. Let $\tau$ be a type such that for every finite prefix $p$ of $\beta \in \Gamma^\infty$ there exists $q \in \Gamma^*$ with $pq \leq \beta$ and*

- *the $\mathcal{L}(k)$-factorization $G$ of $pq$ has type $\tau$, and*
- *$pq = w_0 v_1 w_1 \cdots v_\ell w_\ell$ for some $v_i \in \Gamma^*$ such that the $w_i$'s cover the positions of $G$.*

*Then $s \leq_\mathcal{R} t$ for all linked pairs $(s, e)$ and $(t, f)$ of $M$ with $\alpha \in [s][e]^\omega$ and $\beta \in [t][f]^\omega$.*

*Proof:* Suppose $\alpha \in [s][e]^\omega$ and $\beta \in [t][f]^\omega$. We can write $\beta \in p[f]^\omega$ with $h(p) = t$. By assumption, there exists $q \in \Gamma^*$ such that $pq$ is a prefix of $\beta$ with $\mathcal{L}(k)$-factorization $G$ of type $\tau$. Moreover, we have a factorization $pq = w_0 v_1 w_1 \cdots v_\ell w_\ell$ such that the positions of $G$ are covered by the $w_i$'s. Let $r = h(pq)$. We have $r \leq_\mathcal{R} t$ because $p$ is a prefix of $pq$. By Lemma 11 we have $h(w_0 u_1 w_1 \cdots u_\ell w_\ell) = h(w_0 v_1 w_1 \cdots v_\ell w_\ell)$. Since we can write $\alpha \in w[e]^\omega$ such that $h(w) = s$ and $w_0 u_1 w_1 \cdots u_\ell w_\ell$ is a prefix of $w$, we conclude $s \leq_\mathcal{R} r \leq_\mathcal{R} t$. □

Let $G = (y_1, v_1, \ldots, y_m, v_m)$ be a factorization. A factorization $F = (x_1, u_1, \ldots, x_\ell, u_\ell)$ is a *subfactorization* of $G$, denoted by $F \preceq G$, if for every $i \in \{1, \ldots, \ell\}$ there exists $j \in \{1, \ldots, m\}$ such that $v_j = p u_i q$ and $x_i = y_j + |p|$ for some $p, q \in \Gamma^*$. Intuitively, this means that every $u_i$ is covered by some $v_j$. Let $G$ and $G'$ be factorizations of the same type. Then, there is a one-to-one correspondence between the positions of $G$ and the positions of $G'$. Hence, every subfactorization $F \preceq G$ induces a subfactorization $F' \preceq G'$.

For every factorization $F = (x_1, u_1, \ldots, x_\ell, u_\ell)$ with $x_1 = 1$ we define the monomial $P_F = u_1 \Gamma^* u_2 \cdots \Gamma^* u_\ell \Gamma^\infty$ of degree $|u_1 \cdots u_\ell|$. Now, whenever $F$ is a factorization of a word $\alpha$, then $\alpha \in P_F$. The converse does not hold, but if $\alpha \in P_F$, then there exists a factorization $F'$ of $\alpha$ with type $(u_1, \ldots, u_\ell)$. Next, we give a canonical way of turning a membership $\alpha \in P_F$ into such a factorization $F'$.

Let $P = u_1\Gamma^* u_2 \cdots \Gamma^* u_\ell \Gamma^\infty$ be a monomial and suppose $\alpha \in P$. Write $\alpha = u_1 s_1 u_2 \cdots s_{\ell-1} u_\ell \beta$ such that $(|s_1|, \ldots, |s_{\ell-1}|)$ is minimal in the lexicographic order, i.e., we first minimize $|s_1|$, then $|s_2|$, and so on. We can think of this as greedily minimizing the lengths of the $s_i$'s one after another. Now, the *greedy factorization* for $\alpha \in P$ is $F' = (x_1, u_1, \ldots, x_\ell, u_\ell)$ with $x_i = 1 + |u_1 s_1 u_2 \cdots s_{i-1}|$.

**Proposition 13.** *Let $h : \Gamma^* \to M$ be a homomorphism with $h(\Gamma^+) \in \mathbf{B_1}$ and let $\alpha \in [s][e]^\omega$ and $\beta \in [t][f]^\omega$ for some linked pairs $(s, e)$ and $(t, f)$ of $M$. If $\alpha$ and $\beta$ are contained in the same monomials $w_1\Gamma^* w_2 \cdots \Gamma^* w_n \Gamma^\infty$ of degree at most $4|M|^2$, then $s\,\mathcal{R}\,t$.*

*Proof:* Let $k = |M|$. We shall first give an intuitive outline of our proof. We consider the $\mathcal{R}(k)$-factorization $F$ of $\alpha$. This converts to a factorization $F'$ of $\beta$. Then we choose a prefix $pq$ of $\beta$ such that its $\mathcal{L}(k)$-factorization $G'$ is "as far to the right as possible" in a certain sense. Next, the factorization $G'$ of $\beta$ is converted into a factorization $G$ of $\alpha$. This process makes use of the factorization $F'$ to ensure that on $\alpha$ the factorization $G$ is sufficiently far to the right of $F$. Using Proposition 4, the crucial step is to show that $F \vee G$ and $F' \vee G'$ have the same type. This step was inspired by a proof of Klíma [19]. Finally, applying Lemma 12, we obtain $s \leq_{\mathcal{R}} t$. Since the situation is symmetric in $\alpha$ and $\beta$, we conclude $s\,\mathcal{R}\,t$.

Let $F = (x_1, u_1, \ldots, x_\ell, u_\ell)$ be the $\mathcal{R}(k)$-factorization of $\alpha$. Note that $\alpha \in P_F$ and the degree of $P_F$ is at most $2k^2$. Therefore, $\beta \in P_F$ by assumption. Let $F' = (x'_1, u_1, \ldots, x'_\ell, u_\ell)$ be the greedy factorization for $\beta \in P_F$.

There exists a type $\tau$ such that for every prefix $p$ of $\beta$ there is a prefix $pq$ of $\beta$ with an $\mathcal{L}(k)$-factorization of type $\tau$. If $\beta$ is an infinite word, then this means that there are infinitely many such prefixes $pq$ of $\beta$.

Consider some prefix $pq$ of $\beta$ with an $\mathcal{L}(k)$-factorization $G' = (y'_1, v_1 \ldots, y'_m, v_m)$ of type $\tau$ such that $y' > x'$ for as many positions $y'$ of $G'$ and positions $x'$ of $F'$ as possible. Let $H' = F' \vee G'$. We have $\beta \in P_{H'}$ and the degree of $P_{H'}$ is at most $4k^2$. Thus $\alpha \in P_{H'}$. Let $H$ be the greedy factorization for $\alpha \in P_{H'}$. Further, let $G \preceq H$ be the subfactorization of $H$ induced by $G' \preceq H'$. Note that we cannot directly transfer the factorization $G'$ of $\beta$ to the word $\alpha$ because we want that $G = (y_1, v_1, \ldots, y_m, v_m)$ is "sufficiently far to the right". Next, we show $H = F \vee G$.

We claim that for all $i \in \{1, \ldots, \ell\}$, for all $0 \leq j < |u_i|$, and for all $r \in \{1, \ldots, m\}$ we have

$$x_i + j < y_r \text{ iff } x'_i + j < y'_r \text{ and}$$
$$x_i + j \leq y_r \text{ iff } x'_i + j \leq y'_r.$$

Using property "1" of Proposition 4, we see that $F$ is the greedy factorization for $\alpha \in P_F$. Therefore, $x'_i + j < y'_r$ in $\beta$ implies $x_i + j < y_r$ in $\alpha$. Similarly, $x'_i + j \leq y'_r$ in $\beta$ implies $x_i + j \leq y_r$ in $\alpha$. Suppose $x_i + j < y_r$ in $\alpha$. Let

$$J = (x_1, w_1, \ldots, x_i, w_i) \vee (y_r, v_r, \ldots, y_m, v_m).$$

We have $\alpha \in P_J$ and the degree of $P_J$ is at most $4k^2$. Hence, $\beta \in P_J$ and therefore $x'_i + j < y'_r$ by property "2" of Proposition 4 and by choice of $pq$. Suppose $x_i + j \leq y_r$ in $\alpha$. If $x_i + j < y_r$, then we are done by the previous consideration. So suppose $x_i + j = y_r$. We have $\alpha \in P_J$ with $J$ defined as above. Now, $\beta \in P_J$ implies $x'_i + j \leq y'_r$. Note that we cannot conclude $x'_i + j = y'_r$ at this point. This proves the claim.

The above claim shows that indeed $H = F \vee G$. Let $\tilde{p}\tilde{q}$ such that $pq \leq \tilde{p}\tilde{q} \leq \beta$ and $\tilde{p}\tilde{q}$ has an $\mathcal{L}(k)$-factorization of type $\tau$. Then, by property "2" of Proposition 4, the factors of the $\mathcal{L}(k)$-factorization of $\tilde{p}\tilde{q}$ can only lie further to the right than those of the $\mathcal{L}(k)$-factorization of $pq$. Thus considering the $\mathcal{L}(k)$-factorization of $\tilde{p}\tilde{q}$ instead of $pq$ leads to the same factorization $H$ of $\alpha$. Hence, Lemma 12 shows $s \leq_{\mathcal{R}} t$. The situation is symmetric in $\alpha$ and $\beta$. Therefore, $s \mathcal{R} t$. □

We are now ready to prove Theorem 5.

*Proof (Proof of Theorem 5):* "1 ⇒ 2": By Lemma 8, every monomial $w_1\Gamma^*w_2\cdots\Gamma^*w_n\Gamma^\infty$ is definable in $\Sigma_1[<,+1,\min]$. Hence, the Boolean closure of such languages is contained in the Boolean closure of $\Sigma_1[<,+1,\min]$.

"2 ⇒ 3": The condition $\mathrm{Synt}(L) \in \mathbf{B_1}$ is shown in Lemma 9. By Lemma 10, for every linked pair $(s,e)$ of $\mathrm{Synt}^1(L)$ we have $[s] \subseteq L$ if and only if $[s][e]^\omega \subseteq L$. This is equivalent to the condition for linked pairs in "3b", see [10, Proposition 6.4]. The implication "3 ⇒ 4" is trivial since $L$ is recognized by its syntactic homomorphism.

"4 ⇒ 1": We write $\alpha \equiv \beta$ if $\alpha$ and $\beta$ are contained in the same monomials $w_1\Gamma^*w_2\cdots\Gamma^*w_n\Gamma^\infty$ of degree at most $4|M|^2$. Every $\equiv$-class is a finite Boolean combination of such monomials. It therefore suffices to show that $\beta \equiv \alpha \in L$ implies $\beta \in L$. Suppose $\alpha \in [s][e]^\omega \subseteq L$ and $\beta \in [t][f]^\omega$ for some linked pairs $(s,e)$ and $(t,f)$. By Proposition 13 we see that $\alpha \equiv \beta$ implies $s \mathcal{R} t$. Thus $[t][f]^\omega \subseteq L$ and in particular $\beta \in L$. □

## 5 The Fragment $\mathbb{B}\Sigma_1[<,+1,\min,\max]$ over $\Gamma^*$

In this section we give a new self-contained proof of Knast's result for dot-depth one [20]. Another proof was given by Thérien [37]. Both Knast's and Thérien's proof rely on so-called finite categories. Our proof uses only elementary algebraic concepts like Green's relations. The main part of the proof builds on Proposition 13. Note that a language $L \subseteq \Gamma^*$ is definable in $\mathbb{B}\Sigma_1[<,+1,\min,\max]$ over $\Gamma^\infty$ if and only if $L$ is definable in this fragment over $\Gamma^*$.

**Theorem 14.** *Let $L \subseteq \Gamma^*$. The following are equivalent:*

1. *$L$ has dot-depth one, i.e., $L$ is a finite Boolean combination of monomials $w_1\Gamma^*w_2\cdots\Gamma^*w_n$.*
2. *$L$ is definable in $\mathbb{B}\Sigma_1[<,+1,\min,\max]$.*
3. *$\mathrm{Synt}(L) \in \mathbf{B_1}$.*
4. *$L$ is recognized by some homomorphism $h: \Gamma^* \to M$ with $h(\Gamma^+) \in \mathbf{B_1}$.*

*Proof:* "1 ⇒ 2": By Lemma 8, every language of the form $w_1\Gamma^*w_2\cdots\Gamma^*w_n$ is definable in $\Sigma_1[<,+1,\min,\max]$. Hence, the Boolean closure of such languages is contained in the Boolean closure of $\Sigma_1[<,+1,\min,\max]$. "2 ⇒ 3": This is Lemma 9. The implication "3 ⇒ 4" is trivial.

"4 ⇒ 1": We write $u \equiv v$ if $u,v \in \Gamma^*$ are contained in the same monomials $w_1\Gamma^*w_2\cdots\Gamma^*w_n$ of degree at most $4|M|^2$. Every $\equiv$-class is a Boolean combination of such monomials. Thus it suffices to show $h(u) = h(v)$ whenever $u \equiv v$. Applying Proposition 13 with $e = f = 1$ shows $h(u) \mathcal{R} h(v)$ if $u \equiv v$. The *reversal* of a word $w = a_1 \cdots a_n$ with $a_i \in \Gamma$ is $w' = a_n \cdots a_1$. Let $u'$ and $v'$ be the reversals of $u$ and $v$, respectively. Now, $u \equiv v$ implies $u' \equiv v'$. By Proposition 13 we have $h(u') \mathcal{R} h(v')$ in the reversal of $M$. This in turn is equivalent to $h(u) \mathcal{L} h(v)$ in $M$. Thus $h(u) = h(v)$ since $M$ is aperiodic [23, Proposition A.2.9]. Therefore, for every $x \in M$ the language $h^{-1}(x)$ is a Boolean combination of monomials. □

# 6 The Fragment $\mathbb{B}\Sigma_1[<, +1, \min, \max]$ over $\Gamma^\infty$

In this section, we incorporate the max-predicate. This leads to an effective characterization of the first-order fragment $\mathbb{B}\Sigma_1[<, +1, \min, \max]$ over finite and infinite words. The major difference between Theorem 15 below and Theorem 5 is that the "topological" linked pair condition is slightly different. To express this new condition, we have to use the pure syntactic homomorphism which can distinguish between finite and infinite words.

**Theorem 15.** *Let $L \subseteq \Gamma^\infty$ be regular. The following assertions are equivalent:*
1. *$L$ is a finite Boolean combination of monomials $w_1\Gamma^*w_2\cdots\Gamma^*w_n\Gamma^\infty$ and $w_1\Gamma^*w_2\cdots\Gamma^*w_n$.*
2. *$L$ is definable in $\mathbb{B}\Sigma_1[<, +1, \min, \max]$.*
3. *The pure syntactic homomorphism $h_+ : \Gamma^* \to \mathrm{Synt}_+(L)$ satisfies*
    a) $\mathrm{Synt}(L) \in \mathbf{B_1}$, *and*
    b) *for all linked pairs $(s,e)$ and $(t,f)$ of $\mathrm{Synt}_+(L)$ with $s \mathcal{R} t$ and $e \neq 1 \neq f$ we have $[s][e]^\omega \subseteq L \Leftrightarrow [t][f]^\omega \subseteq L$.*
4. *$L$ is recognized by a homomorphism $h : \Gamma^* \to M$ with $h(u) = 1$ only if $u = 1$ satisfying*
    a) $h(\Gamma^+) \in \mathbf{B_1}$, *and*
    b) *for all linked pairs $(s,e)$ and $(t,f)$ of $M$ with $s \mathcal{R} t$ and $e \neq 1 \neq f$ we have $[s][e]^\omega \subseteq L \Leftrightarrow [t][f]^\omega \subseteq L$.*

Before proving Theorem 15 at the end of this section, we give a counterpart of Lemma 10 for infinite words.

**Lemma 16.** *Let $L \subseteq \Gamma^\infty$ be definable in $\mathbb{B}\Sigma_1[<, +1, \min, \max]$. If $h : \Gamma^* \to M$ is a surjective homomorphism recognizing $L$ such that $h(u) = 1$ only if $u = 1$, then $[s][e]^\omega \subseteq L \Leftrightarrow [t][f]^\omega \subseteq L$ for all linked pairs $(s,e)$ and $(t,f)$ of $M$ with $s \mathcal{R} t$ and $e \neq 1 \neq f$.*

*Proof:* Let $\varphi \in \Sigma_1[<, +1, \min, \max]$ be a sentence. If $\alpha \in \Gamma^\omega$ models $\varphi$, then there is a finite prefix $u$ of $\alpha$ such that for every $\beta \in \Gamma^\omega$ we have $u\beta \models \varphi$. This is because $\alpha \models \varphi$ yields some satisfying assignment for the variables, and positions beyond the last position of this assignment have no influence.

Let $L$ be defined by a formula with quantifier depth $d$, let $t = sx$ and $s = ty$ for $x, y \in M$. Consider $\alpha_1 = \hat{s}\hat{e}^\omega$ for $\hat{s} \in [s]$ and $\hat{e} \in [e]$, and let $\hat{x} \in [x]$, $\hat{y} \in [y]$, and $\hat{f} \in [f]$. By the above consideration, there exists a finite prefix $u = \hat{s}\hat{e}^n$ of $\alpha_1$ such that $\beta_1 = u\hat{x}\hat{f}^\omega$ models at least the same formulas in $\Sigma_1[<, +1, \min, \max]$ with quantifier depth at most $d$ as $\alpha_1$ does. Similarly, there exists a prefix $v = u\hat{x}\hat{f}^m$ of $\beta_1$ such that $\alpha_2 = v\hat{y}\hat{e}^\omega$ models at least the same formulas in $\Sigma_1[<, +1, \min, \max]$ with quantifier depth at most $d$ as $\beta_1$ does. We continue this process and construct $\alpha_1, \beta_1, \alpha_2, \beta_2, \ldots$ such that each word satisfies at least the same formulas with quantifier depth $d$ as its predecessor. There are only finitely many nonequivalent $\Sigma_1[<, +1, \min, \max]$-formulas with quantifier depth at most $d$. Hence, there exist words $\alpha_i \in [s]\hat{e}^\omega$ and $\beta_i \in [t]\hat{f}^\omega$ which satisfy the same formulas in $\Sigma_1[<, +1, \min, \max]$ with quantifier depth at most $d$. Now, $\alpha_i \in L$ if and only if $\beta_i \in L$. This yields $[s][e]^\omega \subseteq L$ if and only if $[t][f]^\omega \subseteq L$. □

Combining Theorem 5, Theorem 14, and Lemma 16 yields the following proof of Theorem 15.
*Proof (Proof of Theorem 15):* "1 ⇒ 2": By Lemma 8, every monomial $w_0\Gamma^*w_1\cdots\Gamma^*w_n\Gamma^\infty$ or $w_0\Gamma^*w_1\cdots\Gamma^*w_n$ is definable in $\Sigma_1[<, +1, \min, \max]$. Therefore, the Boolean closure of such languages is contained in $\mathbb{B}\Sigma_1[<, +1, \min, \max]$.

"2 ⇒ 3": Let $L$ be defined by $\varphi \in \mathbb{B}\Sigma_1[<,+1,\min,\max]$. Lemma 9 shows $\mathrm{Synt}(L) \in \mathbf{B_1}$. The condition "3b" for the linked pairs follows from Lemma 16.

"3 ⇒ 4": This is trivial, since $h_+ : \Gamma^* \to \mathrm{Synt}_+(L)$ recognizes $L$ and $h_+$ maps only the empty word to $1 \in \mathrm{Synt}_+(L)$.

"4 ⇒ 1": Consider $L_\omega = L \cap \Gamma^\omega$ and let $L_\infty$ be the union of the language $L_\omega$ and the following language over finite words:

$$\bigcup \{[s] \mid L_\omega \cap [s][e]^\omega \neq \emptyset \text{ for some linked pair } (s,e) \text{ of } M\}.$$

Then $L_\infty$ satisfies condition "4" in Theorem 5 for the homomorphism $h$ and hence $L_\infty$ is a finite Boolean combination of monomials $w_1\Gamma^* w_2 \cdots \Gamma^* w_n \Gamma^\infty$. Since $\Gamma^\omega$ is a finite Boolean combination of languages $\Gamma^* a$ and $a\Gamma^\infty$ for $a \in \Gamma$, we see that $L_\omega = L_\infty \cap \Gamma^\omega$ is a finite Boolean combination of monomials $w_1\Gamma^* w_2 \cdots \Gamma^* w_n \Gamma^\infty$ and $w_1\Gamma^* w_2 \cdots \Gamma^* w_n$. Consider $L_* = L \cap \Gamma^*$. Lemma 1 shows that $L_*$ is recognized by $h$. Therefore, $L$ is a finite Boolean combination of monomials $w_0\Gamma^* w_1 \cdots \Gamma^* w_n$ by Theorem 14. Thus $L = L_* \cup L_\omega$ is of the required form. □

## 7 The Fragment $\mathbb{B}\Sigma_1[<,+1,\min]$ over $\Gamma^\omega$

If we consider infinite words only, the predicate max is always false. Hence, the first-order fragments $\mathbb{B}\Sigma_1[<,+1,\min,\max]$ and $\mathbb{B}\Sigma_1[<,+1,\min]$ coincide. In this section we give an effective characterization of this fragment for infinite words. It is a rather straightforward consequence of Theorem 15.

**Theorem 17.** *Let $L \subseteq \Gamma^\omega$ be regular. The following assertions are equivalent:*

1. *$L$ has dot-depth one, i.e., $L$ is a finite Boolean combination of monomials of the form $w_1\Gamma^* w_2 \cdots \Gamma^* w_n \Gamma^\omega$.*
2. *$L$ is definable in $\mathbb{B}\Sigma_1[<,+1,\min]$ over $\Gamma^\omega$.*
3. *The pure syntactic homomorphism $h_+ : \Gamma^* \to \mathrm{Synt}_+(L)$ satisfies*
   a) *$\mathrm{Synt}(L) \in \mathbf{B_1}$, and*
   b) *for all linked pairs $(s,e)$ and $(t,f)$ of $\mathrm{Synt}_+(L)$ with $s \mathrel{\mathcal{R}} t$ and $e \neq 1 \neq f$ we have $[s][e]^\omega \subseteq L \Leftrightarrow [t][f]^\omega \subseteq L$.*
4. *$L$ is recognized by a homomorphism $h : \Gamma^* \to M$ with $h(u) = 1$ only if $u = 1$ satisfying*
   a) *$h(\Gamma^+) \in \mathbf{B_1}$, and*
   b) *for all linked pairs $(s,e)$ and $(t,f)$ of $M$ with $s \mathrel{\mathcal{R}} t$ and $e \neq 1 \neq f$ we have $[s][e]^\omega \subseteq L \Leftrightarrow [t][f]^\omega \subseteq L$.*

*Proof:* "1 ⇒ 2": If $L$ is a Boolean combination of monomials $w_1\Gamma^* w_2 \cdots \Gamma^* w_n \Gamma^\omega$, then $L$ can also be written as a Boolean combination of monomials $w_1\Gamma^* w_2 \cdots \Gamma^* w_n \Gamma^\infty$ and $\Gamma^* a$ for $a \in \Gamma$. By Theorem 15 the language $L$ is definable in $\mathbb{B}\Sigma_1[<,+1,\min,\max]$ over $\Gamma^\infty$. Since max is false for all positions of an infinite word, $L$ is definable in $\mathbb{B}\Sigma_1[<,+1,\min]$ over $\Gamma^\omega$.

"2 ⇒ 3": Let $L$ be definable in the fragment $\mathbb{B}\Sigma_1[<,+1,\min]$ over $\Gamma^\omega$. Then $L$ is definable in $\mathbb{B}\Sigma_1[<,+1,\min,\max]$ over $\Gamma^\infty$ and by Theorem 15 the claim follows. "3 ⇒ 4": Trivial.

"4 ⇒ 1": Let $L_\infty$ be the union of $L$ and the following language over finite words

$$\bigcup \{[s] \mid L \cap [s][e]^\omega \neq \emptyset \text{ for some linked pair } (s,e) \text{ of } M\}.$$

Now, $L_\infty$ satisfies condition "4" in Theorem 5 for the homomorphism $h$ and we obtain that $L_\infty$ is a Boolean combination of monomials $w_1\Gamma^* w_2 \cdots \Gamma^* w_n \Gamma^\infty$. Moreover, $L = L_\infty \cap \Gamma^\omega$ and $L$ is a Boolean combination of monomials $w_1\Gamma^* w_2 \cdots \Gamma^* w_n \Gamma^\omega$. □

| Fragment | Models | Languages | Algebra + Linked Pairs | |
|---|---|---|---|---|
| $\mathbb{B}\Sigma_1[<,+1,\min]$ | $\Gamma^\infty$ | $\mathbb{B}\{w_1\Gamma^*\cdots\Gamma^*w_n\Gamma^\infty\}$ | $\mathbf{B_1}$ + $\mathcal{R}$-closed | Thm. 5 |
| $\mathbb{B}\Sigma_1[<,+1,\min,\max]$ | $\Gamma^\infty$ | $\mathbb{B}\begin{Bmatrix}w_1\Gamma^*\cdots\Gamma^*w_n\Gamma^\infty\\w_1\Gamma^*\cdots\Gamma^*w_n\end{Bmatrix}$ | $\mathbf{B_1}$ + $\mathcal{R}^+$-closed | Thm. 15 |
| $\mathbb{B}\Sigma_1[<,+1,\min,\max]$ | $\Gamma^*$ | $\mathbb{B}\{w_1\Gamma^*\cdots\Gamma^*w_n\}$ | $\mathbf{B_1}$ | [20], Thm. 14 |
| $\mathbb{B}\Sigma_1[<,+1,\min]$ | $\Gamma^\omega$ | $\mathbb{B}\{w_1\Gamma^*\cdots w_n\Gamma^\omega\}$ | $\mathbf{B_1}$ + $\mathcal{R}^+$-closed | Thm. 17 |

Table 2: Characterizations of the fragment $\mathbb{B}\Sigma_1$ for various signatures and models

Since condition "3" in Theorem 17 is decidable, we obtain the following corollary.

**Corollary 18.** *It is decidable whether a regular language $L \subseteq \Gamma^\omega$ has dot-depth one.* □

**Remark 19.** *Another algebraic framework for infinite words are $\omega$-semigroups [23]. An $\omega$-semigroup $(S_+, S_\omega)$ has two components. The first component $S_+$ is a semigroup equipped with an infinite product operation and $S_\omega$ is the set of results of infinite products. The conditions "3" in Theorem 15 and "3" in Theorem 17 are equivalent to saying that the syntactic $\omega$-semigroup $(S_+, S_\omega)$ satisfies $S_+ \in \mathbf{B_1}$ and $(x^\pi y^\pi)^\pi x^\omega = (x^\pi y^\pi)^\pi y^\omega$ in $S_\omega$ for all $x, y \in S_+$, cf. [23, Theorem VI.3.8 (6)]. Here, $x^\pi \in S_+$ denotes the idempotent generated by $x$ and $x^\omega$ is an infinite product. The two components of an $\omega$-semigroup inevitably distinguish between finite nonempty and infinite words. Therefore, $\omega$-semigroups are only suitable for fragments which can distinguish finite from infinite words. In particular, $\mathbb{B}\Sigma_1[<,+1,\min]$ cannot distinguish between finite and infinite words and condition "3" in Theorem 5 is not an equational $\omega$-semigroup condition.*

## 8 Summary

In Table 2 we summarize our results on alternation-free first-order logic $\mathbb{B}\Sigma_1$. We gave classes of languages for which $\mathbb{B}\Sigma_1[<,+1,\min]$ and $\mathbb{B}\Sigma_1[<,+1,\min,\max]$ are expressively complete. Our main results are characterizations of the syntactic homomorphisms of such languages. These characterizations are combinations of algebraic and topological properties. The topological properties are stated in terms of linked pairs.

An entry "$\mathcal{R}$-closed" in the column "Linked Pairs" of Table 2 stands for the equivalence $[s][e]^\omega \subseteq L \Leftrightarrow [t][f]^\omega \subseteq L$ for all linked pairs $(s, e)$ and $(t, f)$ with $s \mathrel{\mathcal{R}} t$ in the syntactic monoid. For "$\mathcal{R}^+$-closed" this equivalence has to hold for the pure syntactic homomorphism and $e \neq 1 \neq f$.

Over $\Gamma^\infty$ there are two variants of the Cantor topology. The first one is defined by the base sets $u\Gamma^\infty$ for $u \in \Gamma^*$, and base sets for the second one are $u\Gamma^\omega$ and $\{u\}$. A regular language is a finite Boolean combination of Cantor sets of the first kind if and only if its syntactic homomorphism is "$\mathcal{R}$-closed". Boolean combinations of Cantor sets of the second kind correspond to "$\mathcal{R}^+$-closed".

In all cases, the combination of the algebraic and the topological properties gives decidability of the membership problem for the respective fragment.

## Acknowledgments

We thank the anonymous referees of the conference version of this paper for their suggestions which helped to improve the presentation.